\documentclass[dvips]{article}
\usepackage{icrctc07}

\title{Confinement of Cosmic Rays  in Dark Matter clumps}
\shorttitle{CR confinement  in DM clumps}

\authors{W. de Boer$^1$, V.Zhukov$^{1,2}$ }
\shortauthors{V.Zhukov and et al}  
\afiliations{
$^1$ Institut f\"ur Experimentelle Kernphysik, Universit\"at Karlsruhe,76128 Karlsruhe, Germany\\
$^2$ Skobeltsyn Institute of Nuclear Physics, Moscow State University, 119992 Moscow, Russia}
\email{zhukov@physik.uni-karlsruhe.de}

\abstract{
Some part of the relic Dark Matter is distributed in small-scale clumps which 
survived structure formation in inflation cosmological scenario.
The annihilation of DM inside these  clumps is  a strong source of stable charged particles
which can have a substantial density near the clump core. The streaming of the annihilation products
from the clump can   enhance   irregularities in the galactic magnetic field.
This  can produce small scale variations in diffusion coefficient 
affecting propagation of Cosmic Rays.
}

\begin{document}
\maketitle

\section{Introduction}
The Cosmic Ray (CR) propagation below $10^{17}$ eV can be  described  as a resonant 
scattering on the magneto hydrodynamic  turbulences (MHD) with the  scale $k$  equal to 
the particle Larmor radius $r_g$ in  the galactic magnetic field $B$,  
$k_r^{-1}\!\sim r_g=pc/ZeB$ \cite{kulsrud69}.
The MHD turbulences can propagate in space as Alfv\'en  waves with the  velocity 
$v_a\!\sim\! B/\sqrt{4\pi\rho_{H}}$, which depends on the interstellar gas density $\rho_{H}$ 
and is in the order of  $10^7 cm/s$. 
The level of MHD turbulences is proportional to the random  component  of the magnetic field  
$\delta B/B$ and  can be  associated with  fluctuations in interstellar medium (ISM) which 
follow a power law  $W(k)\propto k^{\alpha-2}$  in the range of $k=10^{-20}-10^{-8} cm^{-1}$, 
where $\alpha=1/3$  for the Kolmogorov spectrum.
 The  waves  interact with  the CR and  interstellar medium 
and can be enhanced or damped, depending on the energy flow.
In the self consistent approach the  kinetic equation for the spectral density $W(r,k)$ of MHD 
turbulences in spherical coordinates   can be written as \cite{kulsrud69}:
\begin{small}
\begin{equation}\label{wk}
\frac{\partial W}{\partial t}+\frac{1}{r^2}\frac{\partial }{\partial r} v_a r^2 W-\frac{\partial v_a}
{\partial r}\frac{\partial}{\partial k}kW=(G-S)W 
\end{equation}
\end{small}
The G term describes the enhancement of   turbulences
due to streaming of CR particles and  S represents the damping.  The  growth of turbulences
occurs when the CR streaming  velocity $v_s$ is larger than  the Alf\'ven speed $v_a$ \cite{wentzel74}.  
The growth  depends on the gradient of the CR density $f$ which can be obtained
from the steady state solution of the diffusion equation (here without convection and reacceleration)\cite{berezinsky90}:
\begin{small}
\begin{equation}\label{diff}
\frac{1}{r^2}\frac{\partial }{\partial r}D \frac{\partial}{\partial r}r^2 f -
\frac{\partial }{\partial p}\frac{\partial p }{\partial t}f-\sigma v n_H f=-q(p,r) 
\end{equation}
\end{small}
,where $q(r,p)$ is the source term, $\frac{\partial p }{\partial t }$ energy  losses and
$\sigma$ fragmentation cross section. 
The diffusion coefficient $D$  at resonance  is related to $W(k,r)$ as :  
$D(r)\approx v r_g^2 B^2/12 \pi W(k_r,r)$ and a high level of turbulences  corresponds to 
the small coefficient and local confinement of particles. 
Since the enhancement  and  damping strongly depend on the local environment, 
this  opens a possibility for  spatial variations in propagation  parameters
and therefore  CR density. 
Here we consider how the dark matter (DM)
annihilation can introduce such variations and affect the CR  propagation.

\section{Dark Matter annihilation in clumps}
The N-body cosmological simulations and analytical calculations show that in the inflation scenario
the smallest DM structures, or clumps,  originate from   primordial density  fluctuations. 
These primordial  clumps
are partially destroyed during evolution contributing to the bulk DM but 0.001-0.1 of total 
relic  DM can still reside in  clumps, depending on initial conditions  \cite{goerdt07}.
The clump mass distribution follows $n(M)dM\sim M^{-2}$ with the minimum mass 
$M_{cl}^{min} \sim 10^{-8}-10^{-6} M_{\odot}$  defined 
by free streaming of DM particles  after kinetic decoupling \cite{gurevich97}. 
The  local number density distribution  of clumps $n_{cl}$ depends on the bulk density profile and
tidal destruction in the galaxy \cite{dokuchaev03}.
The clumps density profile  is probably cuspy $\rho_{cl}\propto 1/r^{1.5-2.0}$ but is saturated
at some critical density $\rho_{max}$  forming  a dense core $r_c$ \cite{gurevich97}.
Inside the clump, the DM  of  mass $m_{\chi}$ will annihilate producing stable particles: 
protons, antiprotons, positrons, electrons and gamma rays,
which can be observed on top of the ordinary CR fluxes.
The luminosity  of the  clump for an $i$-component   is:
$q_i(r,p)\sim \frac{\langle \sigma v \rangle  Y_i(p)}{m_{\chi}^2} \rho_{max}^2 r_c^3$,
where $Y_i$ is the yield per annihilation. For most of DM candidates the annihilation goes into 
fermions, predominantly quark-antiquarks pair.
For example the mSUGRA neutralino of  $m_{\chi}=100$ GeV annihilating in $b\bar b$
will produce per annihilation at 1 GeV : 
$\sim$3 positrons or electrons, 0.3 protons or antiprotons and 8 gammas, 
the  precise  energy spectrum from quarks hadronization  is well measured  
in accelerator experiments.
The $\langle \sigma v \rangle$ is the thermally averaged annihilation cross section which
can be estimated  from the observed  relic DM density   in time 
of decoupling ($T_{dec}\sim \frac{m_{\chi}}{20}$):
$\langle\sigma v\rangle \approx (\frac{2\cdot 10^{-27} cm^3 s^{-1}}{\Omega_\chi h^2})$, 
where  $\Omega_\chi h^2 =0.113 \pm 0.009$ \cite{spergel03}. 
Nowadays, at lower temperature,  the cross section  can be the same for the s-waves or only smaller in case of 
p-waves annihilation channel \cite{deboer05pl}.
The DM annihilation (DMA) signal is decreasing fast with the  DM mass,  at least 
as $q\sim m_{\chi}^{-3}$, but the $\rho_{max}$ is not well known and the significant  DMA signal still can be 
obtained even at large  masses of DM particles.
Since the primordial  clumps are much denser than the bulk component,  most of the annihilation signal 
will come from the core of most abundant smallest clumps. The contribution from clumps is usually expressed as
a 'boost factor'  $b \sim \frac{F_{clump}}{F_{total}}$ and $b>\!>1$, probably except galactic center for the cusped bulk propfile.
Taking a clump with $M_{cl}=10^{-8} M_{\odot}$ and  100 AU size  with the average density of 100 GeV/cm$^3$, 
the total yield  for GeV  charged particles will be $\sim 10^{23} s^{-1}$ for $m_{\chi}=100$ GeV. 
Assuming the isothermal  distribution
of clumps  in the galactic halo  of 20 kpc diameter and   normalizing at  $n_{cl}(8.5kpc)=10pc^{-3}$, 
the total luminosity from the DM annihilation in GeV range  in $100$ years will be $\sim10^{45}$ particles, 
to be compared with the SNR explosion delivering $\sim 10^{51}$ particles in the galactic 
disk. The DM clumps  is  a compact and  constant  source of CR and
despite of  smaller luminosity  the local density of produced particles can  exceed the galactic average  
$\langle \rho_{cr} \rangle\sim 10^{-10} cm^{-3}$ producing a gradient in CR density distribution.

\section{MHD turbulences initiated by DM annihilation}
\begin{figure}[h]
  \begin{center}
    \includegraphics*[width=0.5\textwidth,angle=0,clip]{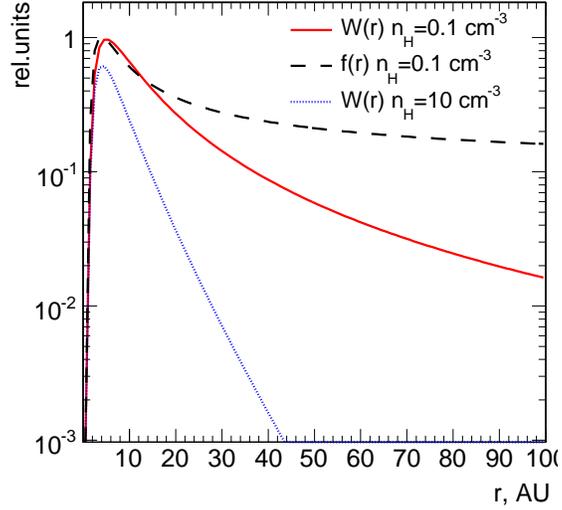}
   \end{center}
   \caption{ The normalized  $W(r)$  and  the  $f(r)$   density  of  GeV protons  in the DM clump 
($\rho_{max}\sim 10^3 GeV/cm^3, m_{\chi}=100 GeV,  v_a \sim 10^6 cm/s,   B\sim 1 \mu G$)}\label{dmclump}
\end{figure}

\begin{figure}[h]
  \begin{center}
    \includegraphics*[width=0.5\textwidth,angle=0,clip]{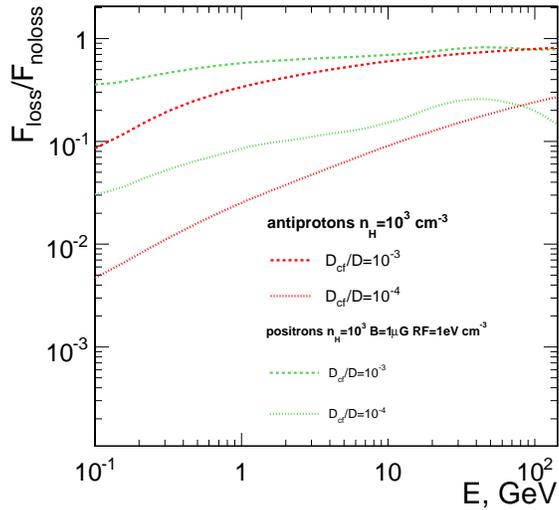}
   \end{center}
   \caption{ Modified by energy looses spectra of antiprotons and positrons from a clump($r_{cf}=10^{-2}pc$)
at different  $D_{cl}/D_{ext}$.}\label{clumpspectra}
\end{figure}

The streaming of charged DMA products  from the  clump   with the drift velocity above  Alfv\'en speed 
 can  increase the level of  local MHD turbulences. 
The amplification of MHD waves parallel to magnetic field  lines can be calculated as \cite{wentzel74}: \\
$
G(r,k)\!\approx\!\frac{\pi^2 e^2 v_a}{k c^2}\int\!\!\int dpd\mu 
 v p^2(1-\mu^2) \delta(p|\mu|-\frac{eB}{kc}) 
\times (\frac{\partial f} {\partial \mu}+
\frac{v_a p}{v} \frac{\partial f}{\partial p}) 
$\\
,where  $\mu$ is the cosine  of scattering angle and 
$\frac{\partial f}{\partial \mu} \sim  \frac{p^2 c^2 }{4 \pi^2 e^2 W } \frac{\partial f}{\partial r}$ 
is the anisotropic term of the CR density distribution which can be obtained from integration  
of diffusion equation (\ref{diff}).
The transverse  waves  are averaged out along propagation path and scattering is not efficient, 
that is, only turbulences along local field lines will be effectively  amplified by streaming.
The growth is reduced in  dense molecular clouds with $n_{H}\sim 10-10^{5} cm^{-3}$
due to ions-neutral  friction which  dissipates energy as 
$S_{H}\sim \frac{1}{2} \langle \sigma_{col} v_a \rangle n_H$, where  
$\langle \sigma_{col} v \rangle\sim 10^{-9} cm^3s^{-1}$  is the collisional  cross section.
The collision of opposite waves leads to a nonlinear damping proportional to the level of turbulences
$S_{nl}\approx\frac{4 \pi v_{t} W}{B^2 r_g^2}=S^0_{nl}W$, where $v_t$ is the gas thermal velocity 
\cite{berezinsky90}.
In the absence of external source of MHD waves the simplified solution of equation 
(\ref{wk})  and (\ref{diff})
neglecting  energy dependencies,  can be calculated for the spatial component analytically:
\begin{small}
\begin{equation}\label{sol}
W(r)\sim \frac{exp(-g/r -s_{H}r)}{r^2 (C_0+s_{nl}exp(-g/r-s_{H}r)/g)} 
\end{equation}
\end{small}
,where $g\!\sim\!\langle \sigma v\rangle Y_{tot}\frac{\rho_{max}^2}{c m_{\chi}^2}$ is related to the 
total clump liminosity, 
$s_H=S_{H}/v_a$,  $s_{nl}=S^0_{nl}/v_a\!<<s_H$ are the dampings,  and $C_0$ is a normalization factor.
In the steady state it will result in fast increase  of $W(r)$ near the clump core  followed by  
an exponential decrease due to damping,  see Figure ~\ref{dmclump}. 
The CR density distribution, obtained  from solution of equation (\ref{diff}) with the diffusion defined by 
(\ref{sol}), will follow the $W(r)$ dependency.
The source function for spectral density $W(k)$ is  cutoff at $k\sim \frac{eB}{m_{\chi}}$ but can be 
extended to smaller  wave numbers by collisions,  reproducing Kolmogorov  spectrum.
More detailed consideration would include energy losses, fragmentation and 
interplay between electrons and protons streaming which is out of the scope of present paper.
For the stable growth the  velocity  of the clump proper motion and the   
convection speed should be below  local $v_a$.
In this case the anisotropic streaming of DMA products  can create a region around the clump with 
small diffusion $D_{cl}\!<\! D_{ext}$, aligned with  the local  $B$ field and  with  size defined by the 
DM density in the clump,  annihilation yields and dumping rates.
The growth will be strongly suppressed inside  dense gas clouds  although particles 
produced in the cusp  of the clump or on the border of the cloud still can be confined. 
The  energy losses and fragmentaion in the confinement region  $r_{cf}$ can modify spectra of
annihilation products. The time scale for the  energy losses $\tau_{loss}(E)$ has to be 
compared with the confinemnt time $\tau_{cf}\!\sim\! r_{cf}^2/6 D_{cl}(E)$. The DMA antiprotons(protons)  will
lose energy in the  gas clouds by ionization below 100 MeV and by nuclear interactions at higher energies.
For positrons(electrons) the synchrotron and inverse Compton are the dominant losses at high and
the  bremsstrahlung at low energies.
The modification of the energy spectra from the  clump due to losses  is shown  in Figure  ~\ref{clumpspectra}
for different  diffusion coefficients.
Thus the  contribution of charged DMA products to the locally observed CR fluxes will depend upon
local environments and location of nearest clumps. 

\begin{figure}[h]
  \begin{center}
    \includegraphics*[width=0.5\textwidth,angle=0,clip]{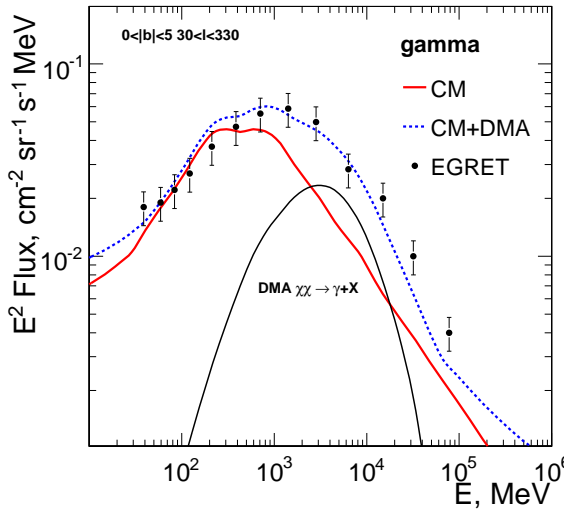}
   \end{center}
   \caption{ Diffusive spectrum of gamma rays from CR in conventional model(CM) \cite{ms04} and 
the  contribution from DMA ($m_{\chi}=70 GeV$).}\label{gamma}
\end{figure}

Gamma rays from the DM clump will have two components, see Figure \ref{gamma}  where
gamma rays have been  calculated with the GALPROP code \cite{ms04}  modified to include DMA 
and small scale variations in propagation parameters. 
First, the direct gamma rays from DMA 
$\chi\chi\rightarrow\gamma+X$ will  produce a bump in the spectrum  at $E\sim 0.1 m_{\chi}$
 independent on the confinement \cite{deboer05pl} .
Second, the electrons(positrons) will contribute to  the lower energy gamma rays via 
 bremsstrahlung and inverse Compton,  and  protons(antiprotons)
to the $\pi^0$ peak  but with somewhat harder spectrum as compare with the gamma rays from CR.
The synchrotron radiation of electrons and positrons will also contribute to the radio waves.
This can be an important source of radio emission in the galactic halo where the  MHD enhancement 
can result  in $\delta B/B>1$.
The secondary contributions will strongly depend on the MHD enhancement and environments:
gas clouds, low density regions in the galactic disk or the  galactic halo.
The spectral features of the gamma  DMA signal can be distinguished from the 
diffusive gamma rays produced by  ordinary CR \cite{deboer05}.
\par
The confinement regions will also affect propagation of the galactic CR. The  low energy external  CR will not 
penetrate deep inside such a clump, their contribution will be suppressed while the  contribution of 
DMA particles  produced in the clump will be enhanced.

\section{Conclusion}

The streaming of  charged particles from DM annihilation in the cuspy  DM clump can locally
enhance level of MHD turbulences reducing the local  diffusion coefficient by orders of magnitude.
The size of confinement
region depends on the luminosity of the clump and damping of turbulences in dense gas clouds.
The confinement regions  will contribute to the small scale variations of propagation parameters and 
therefore CR density.
The gamma rays from the DM annihilation in the
clump can be observed as a point like source with a particular spectrum with the bump from the direct gamma 
production in DM annihilation  and secondary gamma rays produced by the confined charged particles in the clump.
The annihilation in DM clumps will also produce synchrotron radiation even far away from the galactic disk.
The contribution of  charged  DMA products to the observed CR fluxes will  depend on the local
environment and can be modified by   energy losses.
This can change relations between gamma rays and charged components from DMA.

\section{Acknowledgments}
The authors thanks  H.J. V$\rm\ddot{o}$lk,  V.S. Ptuskin,  V. I. Dokuchaev,  I.V.  Moskalenko, A.W. Strong  and P. Blasi  for  useful  discussions.

\bibliography{zhukov_icrc2007}
\bibliographystyle{plain}
\end{document}